\documentclass[%
reprint, 
amsmath,amssymb,
aps,
prb,
superscriptaddress,
showpacs, showkeys]{revtex4-1}

\bibpunct{[}{]}{,}{n}{}{}	

\let\oldaddcontentsline\addcontentsline

\newcommand{\starttocentries}{\let\addcontentsline\oldaddcontentsline}

\newenvironment{changemargin}[2]{
\begin{list}{}{%
\setlength{\topsep}{0pt}%
\setlength{\leftmargin}{#1}%
\setlength{\rightmargin}{#2}%
\setlength{\listparindent}{\parindent}%
\setlength{\itemindent}{\parindent}%
\setlength{\parsep}{\parskip}%
}%
\item[]}{\end{list}}

\usepackage{amsmath,environ}
\NewEnviron{equationsenviron}{
\begin{equation}\begin{split}
  \BODY
\end{split}\end{equation}
}

\usepackage{mathtools}

\usepackage{bm,array}
\usepackage{booktabs} 
\usepackage{multirow} 
\usepackage{graphicx}
\usepackage[table,xcdraw]{xcolor} 
\usepackage{float} 
\usepackage{textcomp}
\usepackage{afterpage} 
\usepackage{verbatim} 
\usepackage{color} 
\usepackage{soul}
\usepackage{amsmath,amssymb}
\usepackage{graphicx}
\usepackage{dcolumn}
\usepackage{bm}
\usepackage{hyperref}
\hypersetup{bookmarksnumbered, pdfpagemode=UseOutlines, 
colorlinks=true, citecolor=blue, filecolor=blue, linkcolor=blue, urlcolor=blue}

\usepackage{breqn} 
\usepackage{bm}
\usepackage{xr}

\usepackage{booktabs} 
\usepackage{multirow}
\usepackage[table,xcdraw]{xcolor} 
\usepackage{float} 
\usepackage{siunitx} 
\usepackage{textcomp}
\usepackage{afterpage} 
\usepackage{verbatim} 
\usepackage{color} 
\usepackage{soul}
\usepackage{amsmath,amssymb,empheq} 
\usepackage{graphicx}
\usepackage{dcolumn}
\usepackage{bm}
\usepackage{hyperref}
\usepackage{xr}
\hypersetup{bookmarksnumbered, pdfpagemode=UseOutlines, 
colorlinks=true, citecolor=blue, filecolor=blue, linkcolor=blue, urlcolor=blue}

     \makeatletter
\let\cat@comma@active\@empty
\makeatother

\makeatletter
\newcommand\footnoteref[1]{\protected@xdef\@thefnmark{\ref{#1}}\@footnotemark}
\makeatother

\graphicspath{{C:/Mallik_PhD/Dropbox/Data_M_Gurram/FND//Talks_Wed-Sci_Tue-Group_FND/PhD_Thesis_Mallik/Figures/Ch7_paper_CVD_Exf_hBN/}}

\begin{document}

\title{Spin transport in two-layer-CVD-hBN/graphene/hBN heterostructures}
\author{M. Gurram}
\thanks{corresponding author}
\email{m.gurram@rug.nl}
\affiliation{Physics of Nanodevices, Zernike Institute for Advanced Materials, University of Groningen, Groningen, The Netherlands}
\author{S. Omar}
\affiliation{Physics of Nanodevices, Zernike Institute for Advanced Materials, University of Groningen, Groningen, The Netherlands}
\author{S. Zihlmann}
\affiliation{Department of Physics, University of Basel, Basel, Switzerland}
\author{P. Makk}
\affiliation{Department of Physics, University of Basel, Basel, Switzerland}
\author{Q.C. Li}
\affiliation{Center for Nanochemistry (CNC), Department of Materials Science and Engineering, College of Engineering, Peking University, Beijing, P.R. China}
\author{Y.F. Zhang}
\affiliation{Center for Nanochemistry (CNC), Department of Materials Science and Engineering, College of Engineering, Peking University, Beijing, P.R. China}
\author{C. Sch\"{o}nenberger}
\affiliation{Department of Physics, University of Basel, Basel, Switzerland}
\author{B.J. van Wees}
\affiliation{Physics of Nanodevices, Zernike Institute for Advanced Materials, University of Groningen, Groningen, The Netherlands}%

\date{\today}


\begin{abstract}
We study room temperature spin transport in graphene devices encapsulated between a layer-by-layer-stacked two-layer-thick chemical vapour deposition (CVD) grown hexagonal boron nitride (hBN) tunnel barrier, and a few-layer-thick exfoliated-hBN substrate.
We find mobilities and spin-relaxation times comparable to that of SiO$_2$ substrate based graphene devices, and obtain a similar order of magnitude of spin relaxation rates for both the Elliott-Yafet and D'Yakonov-Perel' mechanisms.
The behaviour of ferromagnet/two-layer-CVD-hBN/graphene/hBN contacts ranges from transparent to tunneling due to inhomogeneities in the CVD-hBN barriers. Surprisingly, we find both positive and negative spin polarizations for high-resistance two-layer-CVD-hBN barrier contacts with respect to the low-resistance contacts. 
Furthermore, we find that the differential spin injection polarization of the high-resistance contacts can be modulated by DC bias from -0.3 V to +0.3 V with no change in its sign, while its magnitude increases at higher negative bias. These features mark a distinctive spin injection nature of the two-layer-CVD-hBN compared to the bilayer-exfoliated-hBN tunnel barriers.

\begin{description}
\item[PACS numbers]
\verb+85.75.-d+, \verb+73.22.Pr+, \verb+75.76.+j+, \verb+73.40.Gk+
\end{description}
\end{abstract}

\maketitle

\section{Introduction}

Two-dimensional (2D) van der Waals heterostructures of graphene and hexagonal boron nitride (hBN) have gained a lot of attention for charge\cite{64_Britnell2013_NatCom_hBNGrhBN,76_Wang2013_Sci_1D_PartialPickupTransfer,262_wang2017_MTP_electrical_GrhBN_Rev} and spin\cite{48_Zomer2012_PRB_longdist_GrhBN, 49_Marcos2014_PRL_hBNGrhBN, 208_Pep2015_PRB_24um,92_Roche2015_2D_FlaghsiP_rev} transport studies in a high electronic quality graphene. 
An atomically flat and dangling bonds free hBN dielectric provides a neutral environment to probe the intrinsic transport properties of graphene. 
High-mobility graphene encapsulated between two thick-exfoliated-hBN dielectrics resulted in a large spin relaxation length up to 24 $\mu$m with spin diffusion\cite{208_Pep2015_PRB_24um}, and up to 90 $\mu$m with spin drift\cite{103_Pep2016_NL_80p_drift}. 
However, an efficient injection of spin polarized current into graphene is challenging with the conventional oxide tunnel barriers which suffer from pinholes and inhomogeneous growth\cite{22_Han2010_PRL_Tunnel_1L,23_Tombros2007_Nat}, and result in irreproducible  and low spin injection polarizations\cite{22_Han2010_PRL_Tunnel_1L,31_Jozsa2009_PRB_ControlP_drift}.
Recent progress in exploring different 2D materials revealed that an atomically thin, insulating, and pinhole-free nature of single crystalline hBN makes it a promising tunnel barrier\cite{66_Britnell2012_NL_GrhBNGr} for electrical spin injection and detection in graphene\cite{32_Yamaguchi2013_APE_1LhBN_TB}.

Combining the high-mobility graphene with exfoliated-hBN tunnel barrier resulted in a uniform mobility and spin relaxation length across different regions of the encapsulated graphene\cite{35_Gurram2016_PRB}. 
Furthermore, a fully hBN-encapsulated monolayer-graphene with exfoliated-hBN tunnel barriers showed differential spin polarizations of 1-2\% with monolayer-hBN contacts\cite{32_Yamaguchi2013_APE_1LhBN_TB,35_Gurram2016_PRB,36_Singh2016_APL_nsTs_1L2LhBN,249_gurram2017_NComms_biasInducedP}, up to 100\% with bilayer-hBN contacts\cite{249_gurram2017_NComms_biasInducedP}, and up to 6\% with trilayer-hBN contacts\cite{591_Gurram2017_Unpublished}. Thicknesses more than three layers are not suitable for spin injection\cite{33_Kamalakar2014_SciRep_enhanced_CVDhBN,36_Singh2016_APL_nsTs_1L2LhBN,591_Gurram2017_Unpublished} due to very high tunneling interface resistance.
However, for large-scale spintronics applications, it is important to incorporate large-area chemical vapour deposition (CVD) grown hBN tunnel barriers in spin valves\cite{34_Fu2014_JAP_largeScaleCVDhBN,585_Kamalakar2014_APL_spintronics_Gr_CVDhBN_VdWHS,33_Kamalakar2014_SciRep_enhanced_CVDhBN,19_Kamalakar2016_SciRep_Inversion_CVDhBNgr} and magnetic tunnel junctions\cite{260_dankert2015_NanoRes_hBNTMR,75_Piquemal-Banci2016_APL_hBNMTJ}. Therefore, it is interesting to combine the high-mobility graphene with the efficient CVD-hBN tunnel barriers for spintronics devices.

The potential of CVD-hBN as a tunnel barrier for electrical spin injection into graphene has been recently explored\cite{34_Fu2014_JAP_largeScaleCVDhBN,585_Kamalakar2014_APL_spintronics_Gr_CVDhBN_VdWHS,33_Kamalakar2014_SciRep_enhanced_CVDhBN,19_Kamalakar2016_SciRep_Inversion_CVDhBNgr}.
Electrical injection of spin current using a monolayer-CVD-hBN tunnel barrier is inefficient\cite{34_Fu2014_JAP_largeScaleCVDhBN,33_Kamalakar2014_SciRep_enhanced_CVDhBN,19_Kamalakar2016_SciRep_Inversion_CVDhBNgr} due to its low contact resistance-area product $R_{\text{c}} A$ leading to the spin conductivity mismatch problem\cite{50_Thomas2012_PRB_theory}. This can be overcome by increasing the number of layers which would increase the $R_{\text{c}} A$ value leading to an efficient injection of spin current.
In addition, the spin injection efficiency is expected to be larger for a bilayer hBN barrier than for a single layer hBN barrier\cite{17_Wu2014_PRA_hBNGr}.
However, practically, a controlled and direct growth of bilayer or multilayer($>1$ layer) CVD-hBN is difficult\cite{579_gao2013_ACSNano_Growth_1L2L_CVDhBN}. Therefore, 
for our samples, we prepare a two-layer-CVD-hBN tunnel barrier via layer-by-layer-stacking of two individual monolayers of CVD-hBN.
Note that this two-layer-CVD-hBN is different from the bilayer-CVD-hBN in that the former is layer-by-layer-stacked using two individual monolayers while the latter is as-grown.


Furthermore, previously reported spin transport studies in graphene with CVD-hBN tunnel barriers incorporated a bare SiO$_2$/Si substrate\cite{34_Fu2014_JAP_largeScaleCVDhBN,585_Kamalakar2014_APL_spintronics_Gr_CVDhBN_VdWHS,33_Kamalakar2014_SciRep_enhanced_CVDhBN,19_Kamalakar2016_SciRep_Inversion_CVDhBNgr}.
Even though hBN substrates have not been reported to enhance the spin relaxation times of graphene compared to the SiO$_2$/Si substrate\cite{48_Zomer2012_PRB_longdist_GrhBN}, it can increase the mobility and thus the carrier diffusion. 


Therefore here we combine few-layer exfoliated-hBN as a substrate and two-layer-CVD-hBN as a tunnel barrier to obtain both high mobilities and high spin polarizations. 
The mobility of graphene is surprisingly below 3400 cm$^2$V$^{-1}$s$^{-1}$ and spin relaxation time is lower than 400 ps.
In contrast to the results by Kamalakar \textit{et al}.\cite{19_Kamalakar2016_SciRep_Inversion_CVDhBNgr}, we observe both positive and negative spin polarizations for high-$R_{\text{c}} A$ contacts with respect to the low-$R_{\text{c}} A$ contacts. 

We have a similar system as that of reported by Gurram \textit{et al}.\cite{249_gurram2017_NComms_biasInducedP}, wherein the observed behaviour of bias-dependent differential spin injection polarization $p_{\text{in}}$ is unique to the bilayer nature of the hBN barrier. 
Our system is distinctively different from the exfoliated bilayer hBN tunnel barrier as it consist of two individually stacked CVD hBN monolayers. 
This allows us to investigate if the spin injection efficiency does only depend on the barrier thickness or also on different parameters such as relative crystallographic orientation or quality of the interfaces.
Therefore, we also studied the bias-dependent $p_{\text{in}}$ for high-$R_{\text{c}} A$ contacts, and find that the behaviour of the $p_{\text{in}}$ for two-layer-CVD-hBN is different from bilayer-exfoliated-hBN barrier in two ways. 
First, there is no change in sign of the $p_{\text{in}}$ close to zero bias and the sign does not change within the applied DC bias range of $\pm$0.3 V. 
Second, the magnitude of the $p_{\text{in}}$ increases only at higher negative bias.
Our results show first steps towards promising two-layer-CVD-hBN tunnel barriers but point to the utmost importance of the transfer process.

\section{Device fabrication}
\begin{figure}[!tp] 
\includegraphics[scale=1]{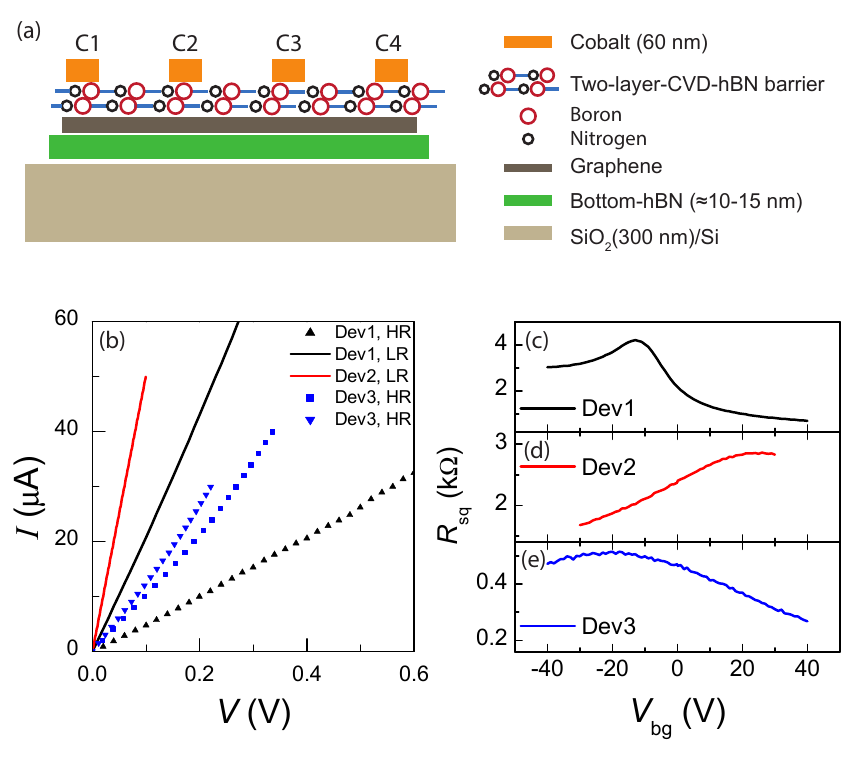}
 \caption{\label{fig:Figure_1} (a) Schematic of the devices prepared with two-layer-CVD-hBN tunnel barriers. Slight displacement in the vertical position of boron and nitrogen atoms of the tunnel barrier represents a crystallographic misalignment between the two CVD-hBN layers. C1-C4 denote the contacts used for the measurements. Other contacts are not shown. (b) Representative three-terminal \textit{I}-\textit{V} curves for three devices, labelled Dev1, Dev2, and Dev3. High-resistance (HR) and low-resistance (LR) contacts are denoted in the legend with symbols and solid-line data, respectively. Within Dev2, all contacts show similar LR behaviour to that of shown here. (c), (d), and (e) show the square resistance $R_{\text{sq}}$ of graphene channel as a function of backgate voltage $V_{\text{bg}}$, for devices Dev1, Dev2, and Dev3, respectively. 
 }
\end{figure}

We have prepared three devices, labelled, Dev1, Dev2, and Dev3. The devices have similar geometry, which is shown in Fig.~\ref{fig:Figure_1}a. CVD-hBN for Dev1 and Dev2, is obtained from Graphene Supermarket Inc., and for Dev3, it is in-house grown by two of the authors\cite{578_song2015_NR_CVDhBN_ChinaGroup}.
Moreover, Dev1 and Dev2 consist of monolayer-exfoliated-graphene, while Dev3 consists of trilayer-exfoliated-graphene. 

The device fabrication is done in two stages. First, the stack of graphene/bottom-hBN on a SiO$_2$/Si substrate is prepared using the dry pick-up and transfer method\cite{44_Zomer2014_APL_FastPickupTransfer}. Then the two-layer-CVD-hBN tunnel barrier is transferred on top of the stack via the conventional wet transfer method\cite{598_lee2010_NL_1stCVDgr_Cu_Transfer}.

In the first stage, we prepared graphene/bottom-hBN stack on a SiO$_2$/Si substrate.
The flakes of graphene and bottom-hBN (typically, $\approx$10-15 nm thick) substrate were exfoliated from highly oriented pyrolytic graphite (HOPG, SPI Supplies, ZYA grade) and hBN-powder (HQ graphene), respectively, on top of a pre-cleaned SiO$_2$(300 nm)/Si substrate using the conventional scotch tape method\cite{232_novoselov2004_Sci_EFeffect_1stpaper}. 
The required flakes were identified via optical microscope and atomic force microscopy.
For making graphene/bottom-hBN stack, we followed the dry pick-up procedure described in Refs.\cite{44_Zomer2014_APL_FastPickupTransfer} and \cite{35_Gurram2016_PRB}. 
In short, we used a glass substrate supporting a polydimethylsiloxane (PDMS) stamp prepared with a polycarbonate (PC) layer to pick up a graphene flake.
Then, the PC/graphene stack is released onto a thick bottom-hBN on a SiO$_2$(300 nm)/Si substrate by melting the PC layer. 
The PC layer is dissolved in chloroform for 5 hours at room temperature.
In order to remove the PC residues from the pick-up and transfer process, the stacks were annealed in Ar/H$_2$ atmosphere at 350 $^{\circ}$C for 12 hours. 

In the second stage, we first prepare the two-layer-CVD-hBN from two individual monolayers of CVD-hBN. 
This is achieved as follows. 
We start with monolayer-CVD-hBN, grown on both sides of a copper (Cu) foil. 
We spin coat PMMA on one side of the Cu foil to protect the CVD-hBN layer and use physical dry etching (O$_2$ plasma) to remove the CVD-hBN on the other side. We then use chemical wet etching to remove the copper by floating the structure PMMA/CVD-hBN/Cu in contact with ammonium persulfate (NH$_4$)$_2$S$_2$O$_4$ etchant solution for 12 hours. While the PMMA/CVD-hBN is still floating, the etchant is replaced with deionized (DI) water several times to clean the contact area of the PMMA/CVD-hBN from the etchant liquid. Then we transfer the cleaned PMMA/CVD-hBN on top of another as-obtained CVD-hBN/Cu/CVD-hBN foil to get the two-layers of CVD-hBN on one side of the Cu foil.

The resulting structure two-layer-CVD-hBN/Cu/CVD-hBN is etched following the same process as before. While the PMMA/two-layer-CVD-hBN is still floating on DI water, we transfer it on to the already prepared graphene/bottom-hBN stack on a SiO$_2$/Si substrate. Then the final stack is put on a hotplate at 180 $^{\circ}$C for two minutes to remove the remaining water. Since the PMMA on top is too thick for lithography, we dissolve it in acetone at 40 $^{\circ}$C for 10 minutes. The resulting device two-layer-CVD-hBN/graphene/bottom-hBN is annealed again in Ar/H$_2$ atmosphere to remove any PMMA residues leftover on the topmost layer.

The electrodes were patterned on the PMMA spin coated stack using electron beam lithography, followed by deposition of ferromagnetic cobalt (Co, 60 nm) capped with aluminum (Al, 5 nm) using electron beam evaporation, and lift-off in acetone at 40 $^{\circ}$C for 10 minutes. A schematic of the final device is depicted in Fig.~\ref{fig:Figure_1}(a). 

Note that the layer-by-layer-stacking of two individual monolayers of CVD-hBN does not guarantee a crystallographic alignment between the monolayers.
The misalignment between the two CVD-hBN layers is schematically represented by a slight displacement in vertical position of the atoms in Fig.~\ref{fig:Figure_1}(a).

\section{Results}
The electrical characterization of the devices is done using a low-frequency lock-in detection technique. All the measurements were carried out at room temperature under vacuum conditions.


The contact resistance of the ferromagnetic tunnelling contacts plays a crucial role in determining its spin injection and detection efficiencies\cite{249_gurram2017_NComms_biasInducedP,19_Kamalakar2016_SciRep_Inversion_CVDhBNgr}.
Therefore, we have characterized the contacts using the three-terminal measurement scheme.
The three-terminal current-voltage (\textit{I}-\textit{V}) characteristics of contacts from three devices are shown in Fig.~\ref{fig:Figure_1}(b).

The differential contact resistance-area product, $R_{\text{c}} A$, of the contacts measured from the three-terminal scheme at zero bias is found to be in the range of 1.0-10.8 k$\Omega\mu$m$^2$. 
In the literature\cite{66_Britnell2012_NL_GrhBNGr,34_Fu2014_JAP_largeScaleCVDhBN,33_Kamalakar2014_SciRep_enhanced_CVDhBN,35_Gurram2016_PRB,249_gurram2017_NComms_biasInducedP}, the reported values of $R_{\text{c}} A$ for monolayer-hBN fall below 4.0 k$\Omega \mu$m$^2$ and for bilayer-hBN, above 4.0 k$\Omega \mu$m$^2$.
Based on these values of $R_{\text{c}} A$, we divide all the contacts of the three devices into two categories, namely, high-resistance (HR, $>$ 4.0 k$\Omega \mu$m$^2$) and low-resistance (LR,  $\leq$ 4.0 k$\Omega \mu$m$^2$) contacts.
Accordingly, Dev1 and Dev3 show contacts ranging from LR to HR, and the Dev2 shows only LR contacts.
LR(HR) contacts of all devices showed linear(non-linear) \textit{I}-\textit{V} behaviour [Fig.~\ref{fig:Figure_1}(b)] which is probably due to the transparent(tunneling) nature of the two-layer-CVD-hBN barriers.

\begin{figure}[!tp]	
 \includegraphics[scale=1]{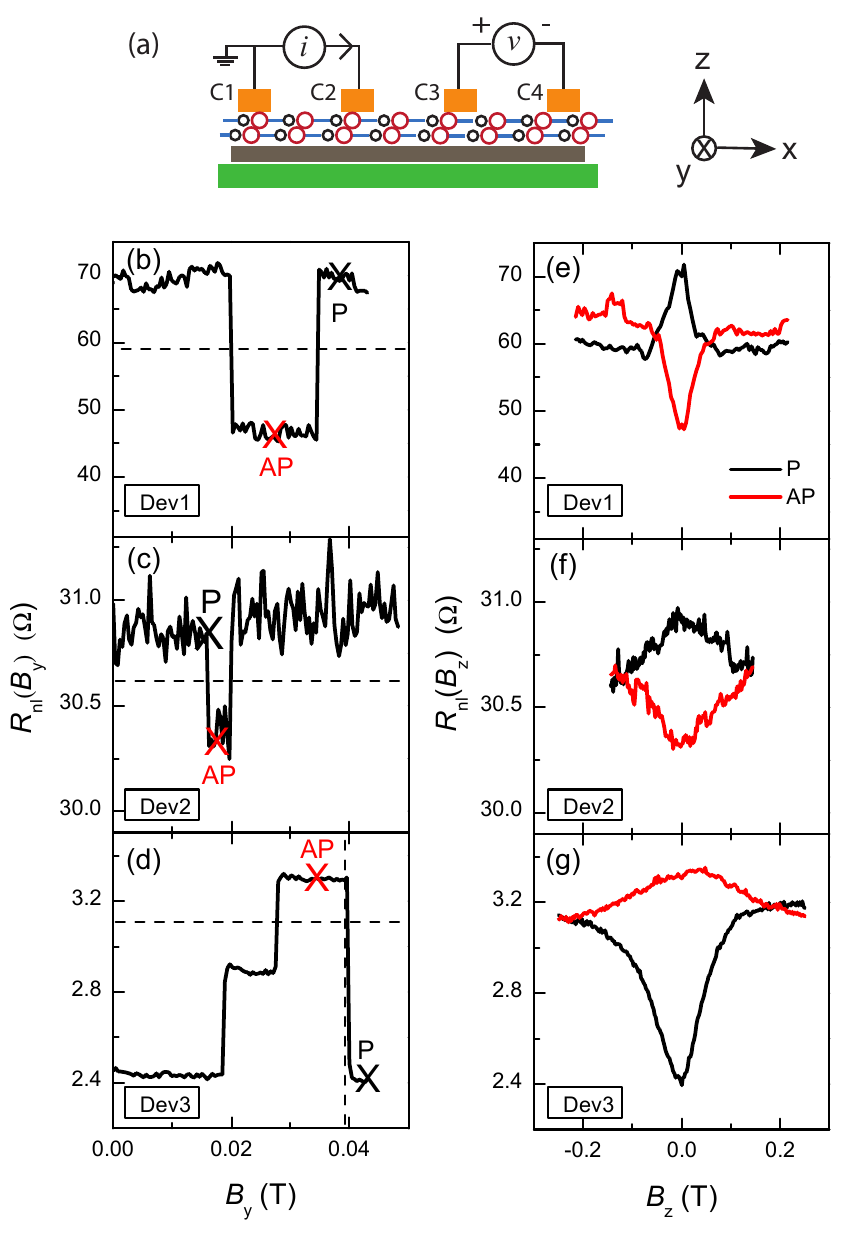}
 \caption{\label{fig:Figure_2} (a) Schematic of the four-terminal non-local measurement geometry for the spin valve and the Hanle measurements. 
(b), (c), and (d) show non-local spin valve signals $R_{\text{nl}}(B_{\text{y}})$ as a function of the magnetic field $B_{\text{y}}$ measured at the carrier densities 0, 1$\times$10$^{12}$, and 4$\times$10$^{12}$ cm$^{-2}$ for the devices Dev1, Dev2, and Dev3, respectively. Horizontal dashed lines represent the background level of the spin valve signal. 
Vertical dashed line in (d) represents the magnetization switching field of the (inner)injector contact. Since the outer-detector contact in Dev3 is also sensitive to the injected spin, we see three switches in its spin valve signal. Parallel (P) and anti-parallel (AP) magnetization configurations of the (inner)injector - (inner)detector contacts pair are denoted by crosses for each spin valve signal.
The Hanle signals $R_{\text{nl}}(B_{\text{z}})$ measured corresponding to the spin valves in (b), (c), and (d), as a function of the magnetic field $B_{\text{z}}$, when the (inner)injector - (inner)detector magnetizations are aligned in P and AP configurations are given in (e), (f), and (g), respectively. 
}
\end{figure}

The spread in the $R_{\text{c}} A$ values could be due to the inhomogeneous growth of CVD-hBN, thickness variation from the wrinkles at the interfaces of two-layer-CVD-hBN/graphene and monolayer-CVD-hBN/monolayer-CVD-hBN during two separate wet transfer processes, and PMMA residues at the interfaces of cobalt/two-layer-CVD-hBN and monolayer-CVD-hBN/monolayer-CVD-hBN.
The low-resistance of the contacts even with two-layers of CVD-hBN can be attributed to the presence of pinholes coming from the inhomogeneous coverage of CVD-hBN, and cracks in CVD-hBN that might be induced during the transfer processes or the annealing step. 


 

We use four-terminal local measurement scheme to characterize the charge transport in graphene where we apply a constant magnitude of AC current $i$ across the outer-electrodes [C1 and C4 in Fig.~\ref{fig:Figure_1}(a)] and measure the voltage drop $v$ across the inner-electrodes (C2 and C3) while sweeping the backgate voltage $V_{\text{bg}}$. Here, the highly p-doped Si is used as a backgate electrode.
The backgate bias $V_{\text{bg}}$ dependence of the square resistance $R_{\text{sq}} = \frac{v}{i}\frac{W}{L}$ of the graphene in three devices is shown in Figs.~\ref{fig:Figure_1}(c)-\ref{fig:Figure_1}(e) where \textit{W} and \textit{L} are width and length of the graphene transport channel. 
Typical values of the $R_{\text{sq}}$ were observed for monolayer graphene in Dev1 and Dev2, whereas a very low-$R_{\text{sq}}$ for Dev3 is due to the trilayer nature of its graphene.
The field-effect mobility of electrons is obtained by fitting the $R_{\text{sq}}$ data using the relation, $R_{\text{sq}} = \frac{1}{n e \mu + \sigma_0} + \rho_{\text{s}}$, with $n$, the carrier density, $e$, the electron charge, $\mu$, the mobility, $\sigma_0$, the residual conductivity, and $\rho_{\text{s}}$, the contribution from short-range scattering\cite{82_zomer2011_APL_mobility_DryTransfer_GrhBN,35_Gurram2016_PRB}. 
The fitting resulted in a surprisingly low electron mobilities $\mu$ = 3400 cm$^2$V$^{-1}$s$^{-1}$ for Dev1, 120 cm$^2$V$^{-1}$s$^{-1}$ for Dev2, and 255 cm$^2$V$^{-1}$s$^{-1}$ for Dev3.
It should be noted that the bottom layers of a few-layer-thick graphene could screen the gate induced electric field. 
However, it was reported that for the multilayer graphene up to five-layers, the bulk carrier density determined from the Hall measurements approximately agrees with the backgate induced carrier density\cite{597_nagashio2010_JJAP_MonoMultiLayer_gr}. 
Therefore, we assume that the obtained value of field-effect mobility of trilayer-graphene in Dev3 is correct.



We use the four-terminal non-local measurement scheme\cite{23_Tombros2007_Nat,35_Gurram2016_PRB} shown in Fig.~\ref{fig:Figure_2}(a) to characterize the spin transport in graphene.
A spin polarized current is injected across a pair of injector contacts [C1 and C2 in Fig.~\ref{fig:Figure_2}(a)] with a constant magnitude of the AC current $i$ = 1 $\mu$A and the diffused spins along the graphene channel are probed as a voltage $v$ across different pair of detector contacts [C3 and C4 in Fig.~\ref{fig:Figure_2}(a)], located outside the charge current path.
The non-local differential spin resistance given by $R_{\text{nl}} = \frac{v}{i}$.

For a clear interpretation of the results presented here, we give $R_{\text{c}} A$ values of the (inner)injector - (inner)detector contacts pair [C2 - C3 in Fig.~\ref{fig:Figure_2}(a)]. Dev1 consists of contacts whose $R_{\text{c}} A$ values are 1.7 k$\Omega \mu$m$^2$-10.8 k$\Omega \mu$m$^2$(LR-HR), Dev2 with 1.2 k$\Omega \mu$m$^2$-1.0 k$\Omega \mu$m$^2$ (LR-LR), and Dev3 has two sets of contacts; set1 with 4.7 k$\Omega \mu$m$^2$-1.4 k$\Omega \mu$m$^2$ (HR-LR), and set2 with 8.6 k$\Omega \mu$m$^2$-2.3 k$\Omega \mu$m$^2$ (HR-LR).

For non-local spin valve measurements, a magnetic field $B_{\text{y}}$ is swept along the easy axes of the Co contacts. Magnetization switching of the contacts at their respective coercive fields results in sharp changes in $R_{\text{nl}}(B_{\text{y}})$ value as shown in Figs.~\ref{fig:Figure_2}(b)-\ref{fig:Figure_2}(d).
The injector-detector pair of Dev1 consisting of LR-HR contacts showed a regular spin valve signal with higher $R_{\text{nl}}$ for parallel (P) and lower $R_{\text{nl}}$ for anti-parallel (AP) configuration of the relative magnetization orientation of the contacts, i.e., spin signal $\Delta R_{\text{nl}} = (R_{\text{nl}}^{\text{P}} - R_{\text{nl}}^{\text{AP}})/2> 0$ [Fig.~\ref{fig:Figure_2}(b)]. 
A similar behaviour is also observed for Dev2 with LR-LR contacts pair [Fig.~\ref{fig:Figure_2}(c)].
Interestingly, Dev3 consisting of HR-LR contacts showed an inverted spin valve signal $\Delta R_{\text{nl}} < 0$ [Fig.~\ref{fig:Figure_2}(d)] whereas, HR-HR and LR-LR combinations of the injector-detector pair resulted in regular spin valve signals $\Delta R_{\text{nl}} > 0$.

\begin{figure}[!tp]	
 \includegraphics[scale=1]{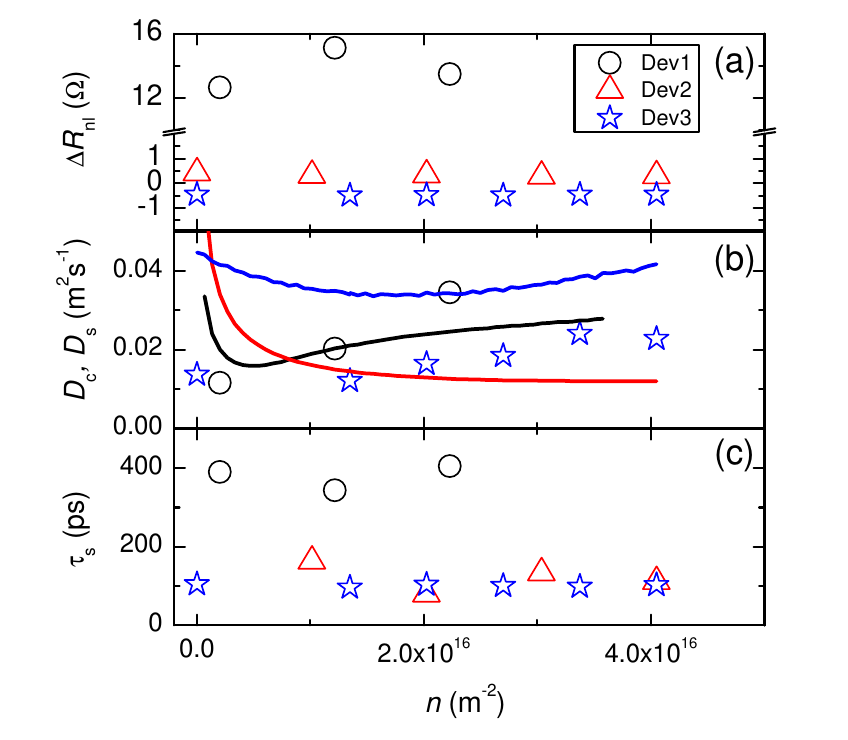}
 \caption{\label{fig:Figure_3} Data extracted from the Hanle spin precession measurements for devices Dev1, Dev2, and Dev3 at different electron carrier densities. 
(a) Non-local Hanle spin precession signal $\Delta R_{\text{nl}} = (R_{\text{nl}}^{\text{P}} - R_{\text{nl}}^{\text{AP}})/2$ at $B_{\text{z}} = 0$. Note that $\Delta R_{\text{nl}}$ for Dev3(for set1 contacts) remains negative for all densities. 
(b) Carrier diffusion constants determined from the charge $D_{\text{c}}$, and the spin $D_{\text{s}}$ transport measurements, lines and symbols, respectively. $D_{\text{s}}$ for Dev2 is not given due to unreliable values obtained from the Hanle fitting. We assume $D_{\text{s}} = D_{\text{c}}$\cite{94_Weber2005_Nat_DcDs} for Dev2 and use $D_{\text{c}}$ values to fit the Hanle data $\Delta R_{\text{nl}}(B_{\text{z}})$, and obtain $\tau_{\text{s}}$.
$D_{\text{c}}$ for Dev3 is calculated from the effective density of states of three-layer graphene\cite{197_maassen2011_PRB_fewLayerGr}. 
(c) Spin relaxation times $\tau_{\text{s}}$. }
\end{figure}

In order to determine the spin transport parameters, we measure non-local Hanle spin precession signals $R_{\text{nl}} (B_{\text{z}})$ for which a magnetic field $(B_{\text{z}})$ is applied perpendicular to the plane of the spin injection, causing the injected spins to precess in plane with a Larmor precession frequency $\omega_{\text{L}} = \frac{g \mu_{\text{B}} B_{\text{z}}}{\hbar}$ where, $g$=2 is the Land\'e factor, $\mu_{\text{B}}$ is the Bohr magneton, and $\hbar$ is the reduced Planck constant.
The Hanle signals $R_{\text{nl}}^{\text{P(AP)}} (B_{\text{z}})$, measured for three devices,
when the relative magnetization orientation of the injector-detector contacts are set in P(AP) configurations are shown in Figs.~\ref{fig:Figure_2}(e)-\ref{fig:Figure_2}(g).
P and AP configurations correspond to the spin valve signals shown in Figs.~\ref{fig:Figure_2}(b)-\ref{fig:Figure_2}(d).
Dev1 and Dev2 showed a regular Hanle signals $R_{\text{nl}} (B_{\text{z}})$ for P (black curve) and AP (red curve) configurations, whereas Dev3 showed an inverted $R_{\text{nl}} (B_{\text{z}})$.

A pure Hanle spin signal $\Delta R_{\text{nl}} (B_{\text{z}})$ is obtained by eliminating the spin-independent signals via $\Delta R_{\text{nl}} = (R_{\text{nl}}^{\text{P}} - R_{\text{nl}}^{\text{AP}})/2$.
We assume a uniform spin injection along the length of the Co/two-layer-CVD-hBN/graphene contacts and fit the $\Delta R_{\text{nl}} (B_{\text{z}})$ data with the one-dimensional steady state solution to the Bloch equation; $D_{\text{s}}\bigtriangledown^{2}\vec{\mu_{\text{s}}} - \vec{\mu_{\text{s}}}/\tau_{\text{s}} + \vec{\omega_{\text{L}}}\times\vec{\mu_{\text{s}}}= 0$ with $\vec{\mu_{\text{s}}}$, the spin accumulation, $D_{\text{s}}$, the spin diffusion constant, and $\tau_{\text{s}}$, the spin relaxation time. 
From the fitting of the Hanle spin signals $\Delta R_{\text{nl}}$ measured at different carrier densities, we obtain the value of $\tau_{\text{s}}$ to be lower than 280 ps for Dev1, 80 ps for Dev2, and 100 ps for Dev3.

In order to study the influence of the LR contacts on spin transport\cite{50_Thomas2012_PRB_theory,117_sosenko2014_Contacts_Ts,116_idzuchi2015_PRB_revisitingTs,120_stecklein2016_PRB_CinducedSpRelxn}, we calculate
the values of ($R_{\text{c}}/R_{\text{s}}$, $L/\lambda_{\text{s}}$) parameters. Here $R_{\text{s}} = R_{\text{sq}}\lambda_{\text{s}}/W$ is the spin resistance of the graphene with $\lambda_{\text{s}} = \sqrt{D_{\text{s}} \tau_{\text{s}}}$, the spin relaxation length, and the ratio $R_{\text{c}}/R_{\text{s}}$ quantifies the back-flow of injected spins into the contacts\cite{50_Thomas2012_PRB_theory}.
For the devices Dev1, Dev2, and Dev3 at different carrier densities we find the values of ($R_{\text{c}}/R_{\text{s}}$, $L/\lambda_{\text{s}}$) in the range of (0.81-12.97, 0.61-1.14), (0.12-3.11, 0.15-2.65), and (13.64-77.81, 0.84-1.19), respectively. 
According to the analysis by Maassen \textit{et al}.\cite{50_Thomas2012_PRB_theory} on contact induced spin relaxation in Hanle spin precession measurements, the low-$R_{\text{c}}/R_{\text{s}}$ values for Dev1 and Dev2 indicate that the spin relaxation in graphene is influenced by spin absorption at the LR contacts and resulted in an underestimated values of the spin transport parameters obtained via Hanle data fitting.
Therefore we estimate the true values of $D_{\text{s}}$ and $\tau_{\text{s}}$ for Dev1 and Dev2 by taking the effect of the low-$R_{\text{c}}/R_{\text{s}}$ contacts into account\cite{50_Thomas2012_PRB_theory}.
For Dev3, high values of $R_{\text{c}}/R_{\text{s}}$ indicates that the spin absorption by contacts is negligible and we can safely assume that the fitted values of $D_{\text{s}}$ and $\tau_{\text{s}}$ represent the true values.
For all devices, the corrected values of $D_{\text{s}}$ and $\tau_{\text{s}}$ are plotted in Figs.~\ref{fig:Figure_3}(b)-\ref{fig:Figure_3}(c) as a function of the electron carrier density. 
For Dev1 and Dev3, we observe a good correspondence between the values of $D_{\text{c}}$ and $D_{\text{s}}$ within a factor of two, confirming the reliability of our analysis\cite{95_Guimaraes2012_NL_suspended,23_Tombros2007_Nat}.
After the correction, the value of $\tau_{\text{s}}$ raised to 400 ps for Dev1, and 160 ps for Dev2. 
Even after the correction, such low value of $\tau_{\text{s}}$ for these devices indicate that the spin relaxation within the graphene channel is dominant.

\section{Discussion}
To prepare our devices using CVD-hBN barriers, we used a similar method as that of Fu \textit{et al}.\cite{34_Fu2014_JAP_largeScaleCVDhBN} and Kamalakar \textit{et al}.\cite{33_Kamalakar2014_SciRep_enhanced_CVDhBN,19_Kamalakar2016_SciRep_Inversion_CVDhBNgr} except, we additionally used a thick-exfoliated-hBN as a substrate. However, despite having the bottom-hBN substrate, we do not observe an enhancement in the mobility of graphene\cite{48_Zomer2012_PRB_longdist_GrhBN}.

From Fig.~\ref{fig:Figure_3}(c), it is clear that even after including the correction from the spin absorption due to the low-$R_{\text{c}}/R_{\text{s}}$ contacts\cite{50_Thomas2012_PRB_theory}, the value of $\tau_{\text{s}}$ is still lower than 400 ps for all three devices.
We do not observe an increased $\tau_{\text{s}}$ in our devices with two-layer-CVD-hBN encapsulating tunnel barriers, compared to the monolayer-CVD-hBN\cite{34_Fu2014_JAP_largeScaleCVDhBN,585_Kamalakar2014_APL_spintronics_Gr_CVDhBN_VdWHS,33_Kamalakar2014_SciRep_enhanced_CVDhBN} encapsulating barriers.
Whereas, in case of exfoliated-hBN encapsulating tunnel barrier, increasing the number of layers from mono to bilayer resulted in an increase of $\tau_{\text{s}}$ due to large $R_{\text{c}} A$ contacts and enhanced screening of polymer contamination by bilayer-hBN\cite{35_Gurram2016_PRB,249_gurram2017_NComms_biasInducedP,36_Singh2016_APL_nsTs_1L2LhBN,591_Gurram2017_Unpublished}.

The lower values of spin relaxation times and mobilities for our hBN-based graphene devices with top CVD-hBN tunnel barrier encapsulation can be attributed to several factors, such as the quality of graphene due to the wet transfer process, the non-uniform CVD-hBN barrier, their improper interface, and the proximity of the lithography residues.
The growth of CVD-hBN can suffer from the inhomogeneous surface coverage, and the copper etching steps could also damage the CVD-hBN and leave some under-etched residues leading to uneven interfacial growth of ferromagnetic cobalt on top\cite{36_Singh2016_APL_nsTs_1L2LhBN} which may cause spin dephasing in graphene via randomly oriented magnetic fringe fields near the contacts\cite{182_dash2011_PRB_InvertedHanle_Al2O3SC}.
Moreover, during the wet transfer of CVD-hBN, some unwanted contamination may get trapped at the interface with graphene, and graphene itself comes in a direct contact with DI water. 
Even though we dry the stack right after the transfer of CVD-hBN on a hot plate, we do not know how many impurities are removed.
Furthermore, we use two-layer (not the as-grown bilayer) CVD-hBN tunnel barrier which may come with additional Cu residues, water molecules, or any hydrocarbon molecules trapped in between the two hBN layers from the preparation steps. During the transfer of one CVD-hBN layer on top of another, even foldings or shrinking of the individual layers can occur.

\begin{figure}[!tp]	
 \includegraphics[scale=1]{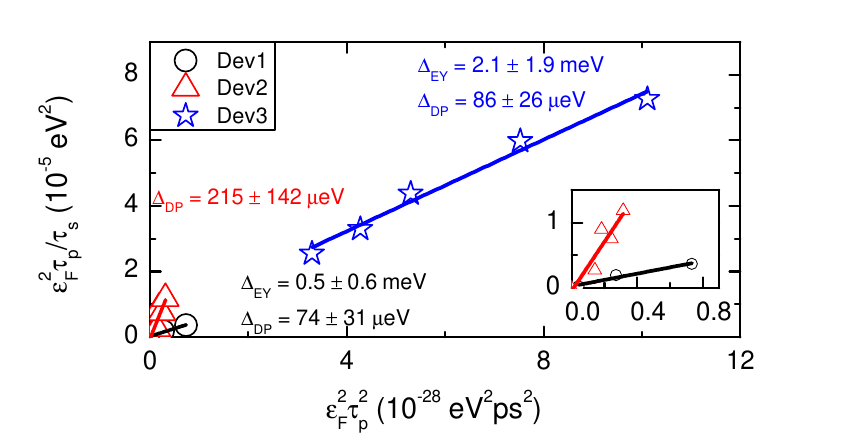}
 \caption{\label{fig:Figure_4} The linear fits (solid lines) of the data using Eq.~(\ref{Eq:EY_DP}) give the spin-orbit coupling strengths of the EY and DP spin relaxation mechanisms, $\Delta_{\text{EY}}$ and $\Delta_{\text{DP}}$, respectively, for three devices. The inset shows the data and fits close to zero. A reliable value of $\Delta_{\text{EY}}$ for Dev2 is not obtained due to non-monotonic relation between $\tau_{\text{s}}$ and $n$\cite{48_Zomer2012_PRB_longdist_GrhBN}[see Fig.~\ref{fig:Figure_3}(c)]. }
\end{figure}

In order to investigate the possible spin relaxation phenomenon causing the low spin relaxation times for graphene in our devices, we analyze the data in Fig.~\ref{fig:Figure_3} by following Zomer \textit{et al}.\cite{48_Zomer2012_PRB_longdist_GrhBN}. We consider Elliott-Yafet (EY) and D'Yakonov-Perel' (DP) mechanisms contributing to the spin relaxation in graphene and analyze the relation between $\tau_{\text{s}}$ and momentum relaxation time, $\tau_{\text{p}}$, using the equation\cite{48_Zomer2012_PRB_longdist_GrhBN},
\begin{equation}	\label{Eq:EY_DP}
\frac{\varepsilon_{\text{F}}^2 \tau_{\text{p}}}{\tau_{\text{s}}} = \Delta_{\text{EY}}^2 + \left( \frac{4 \Delta_{\text{DP}}^2}{\hbar^2}\right) \varepsilon_{\text{F}}^2 \tau_{\text{p}}^2
\end{equation}
where, $\varepsilon_{\text{F}}$ is the Fermi energy of graphene, $\Delta_{\text{EY}}$ and $\Delta_{\text{DP}}$ are the spin-orbit coupling strengths of EY and DP mechanisms, respectively.

The fits to the data for three devices, shown in Fig.~\ref{fig:Figure_4}, using the above equation give $\Delta_{\text{EY}}$ and $\Delta_{\text{DP}}$. 
We calculate the spin relaxation rates due to EY and DP mechanisms from $\tau_{\text{s,EY}}^{-1} = \frac{\Delta_{\text{EY}}^2}{\varepsilon_{\text{F}}^2 \tau_{\text{p}}}$ and $\tau_{\text{s,DP}}^{-1} = \frac{4 \Delta_{\text{DP}}^2 \tau_{\text{p}}}{\hbar^2}$.  The values of ($\tau_{\text{s,EY}}^{-1}$, $\tau_{\text{s,DP}}^{-1}$) for Dev1, Dev2, and Dev3 are found to be in the range of (0.2-2.7, 2.0-2.5) ns$^{-1}$, (-, 10.3-13.8) ns$^{-1}$, and (0.6-1.8, 8.4-9.4) ns$^{-1}$. 
Due to nonlinear nature of the plotted data for Dev2, it cannot be accurately fitted with Eq.~\ref{Eq:EY_DP}.
The relaxation rates for both EY and DP mechanisms are in the similar order of 10$^9$ s$^{-1}$, and a clear dominance of either of the mechanism cannot be distinguished.

From the regular spin valve and Hanle signals for Dev1 [Figs.~\ref{fig:Figure_2}(b) and \ref{fig:Figure_2}(e)], it is evident that the spin polarizations of the LR and the HR contacts have same sign. 
On the contrary, from the inverted spin valve and Hanle signals for Dev3 [Figs.~\ref{fig:Figure_2}(d) and \ref{fig:Figure_2}(g) for set1], at zero DC bias ($V_{\text{in}}$=0V),
we deduce that the spin polarization of the HR contact has an opposite sign with respect to that of the LR contact.



Note that the absolute sign of the spin polarization cannot be determined from the non-local spin transport measurements.
For each device, we assume the polarization of the LR contact to be positive. 
Therefore, for the two-layer-CVD-hBN tunnel barrier contacts, we find both positive and negative spin polarizations for the HR contacts (in the range, 4.7-10.8 k$\Omega \mu$m$^2$) with respect to the LR contacts (in the range, 1.0-2.4 k$\Omega \mu$m$^2$), i.e., there is no consistent correlation between the $R_{\text{c}} A$ values of the HR contacts and their polarization signs (positive or negative). 
This behaviour is different from the resutls reported by Kamalakar \textit{et al}.\cite{19_Kamalakar2016_SciRep_Inversion_CVDhBNgr}, wherein a layer of CVD-hBN tunnel barrier with variable thickness (1-3 layers) is used, and the sign of the spin polarization is reported to be positive only for the contacts with $R_{\text{c}} A \leq 25$ k$\Omega \mu$m$^2$ and negative for $R_{\text{c}} A \geq 170$ k$\Omega \mu$m$^2$.
The authors\cite{19_Kamalakar2016_SciRep_Inversion_CVDhBNgr} used a layer of CVD-hBN with a spatial distribution of thickness varied between 1 and 3 layers. 
Note however, that, this multilayer-CVD-hBN has not been layer-by-layer-stacked but as-grown with inhomogeneous thickness, and the observed behaviour was attributed to the spin-filtering at the cobalt/hBN interface.
Since we do not have a perfect Bernal-stacked bilayer CVD-hBN tunnel barrier, we cannot comment on the possible spin filtering mechanism for the negative polarization of the HR contacts observed here.

In fact, recent study with bilayer-exfoliated-hBN barriers by Gurram \textit{et al}.\cite{249_gurram2017_NComms_biasInducedP} reported that at zero DC bias, different contacts (with $R_{\text{c}} A$ in the range of 4.6-77.1 k$\Omega \mu$m$^2$) showed different signs (positive or negative) of spin polarizations which is also observed here with the two-layer-CVD-hBN barriers.
However, in contrast to the layer-layer-stacked two-layer-CVD-hBN, mechanically exfoliated bilayer-hBN is expected to have crystallographic orientation.
Therefore, it makes more difficult to comment on a possible mechanism causing negative polarization.

\begin{figure}[!tp]	
 \includegraphics[scale=1]{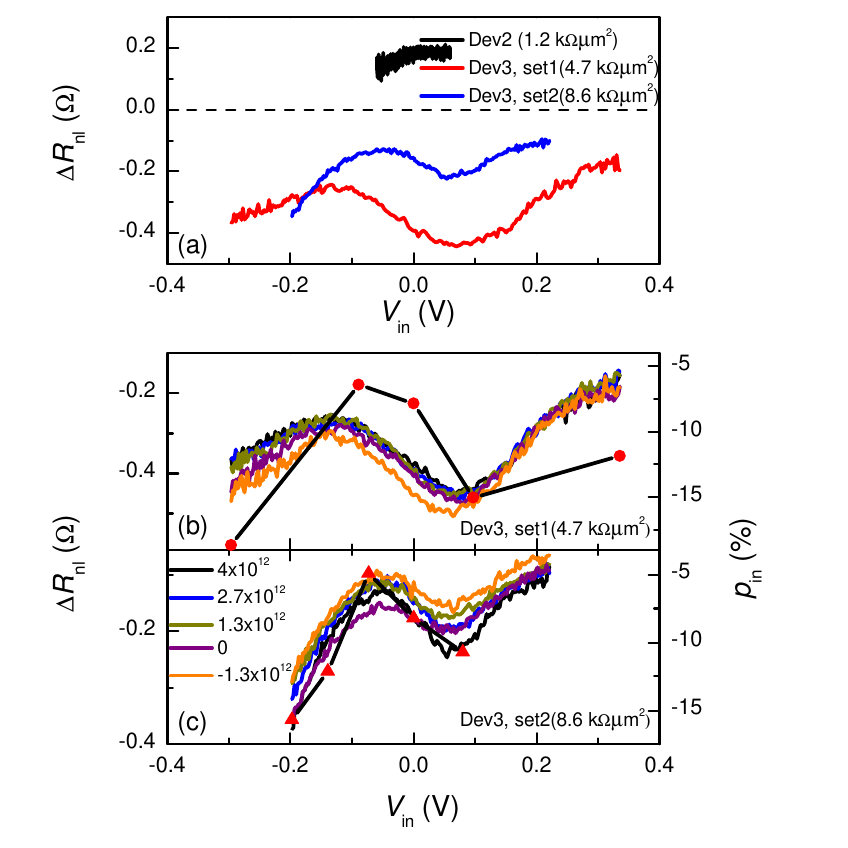}
 \caption{\label{fig:Figure_5} (a) Non-local spin signal $\Delta R_{\text{nl}}$ as a function of the injection bias $V_{\text{in}}$ for Dev2 with LR-LR injector-detector contacts pair, and for Dev3 with two different sets of HR-LR injector-detector contacts pairs. Inversion of the spin signal for Dev3 is due to the inverse polarization of the HR injector contact with respect to the LR detector [see Figs.~\ref{fig:Figure_2}(d) and \ref{fig:Figure_2}(g) for set1]. Dashed line represents $\Delta R_{\text{nl}}=0$. $R_{\text{c}} A$ values of the respective injector contacts, at zero bias, are given in the legend.
The left axis of (b) and (c) shows bias dependent $\Delta R_{\text{nl}}$ for set1 and set2 contacts of Dev3, respectively, at different carrier densities ranging from electrons (n $>$ 0) to holes (n $<$ 0).
Legend in (c) shows the carrier density in cm$^{-2}$.
The right axis of (b) and (c) shows differential spin injection polarization $p_{\text{in}}$ at an electron density of 3.4$\times$10$^{12}$ cm$^{-2}$ for set1 and 4$\times$10$^{12}$ cm$^{-2}$ for set2, respectively. }
\end{figure}

Now we study the bias-dependence of the spin signals $\Delta R_{\text{nl}}$ and differential spin injection polarization $p_{\text{in}}$ of the two-layer-CVD-hBN contacts.
A recent report by Gurram \textit{et al}.\cite{249_gurram2017_NComms_biasInducedP} on the effect of bias applied across the ferromagnetic contacts with a bilayer-exfoliated-hBN barrier revealed a dramatic behaviour of $\Delta R_{\text{nl}}$ and $p_{\text{in}}$ where the sign of the polarization is reversed at a very small bias and its magnitude is increased with bias even up to 100\%. In light of these results, it is interesting to study the bias dependence of the $p_{\text{in}}$ of the two-layer-CVD-hBN barrier contacts. 

In case of application of a bias across a ferromagnetic tunneling contact with transparent regions (i.e., tunnel barrier with pinholes), one would observe an increase(decrease) in the magnitude of spin signal with positive bias for holes(electrons)\cite{31_Jozsa2009_PRB_ControlP_drift} due to a strong local carrier drift in graphene underneath the metallic electrode.
Moreover, the carrier density in graphene underneath such contacts cannot be modified via the back gate voltage as it is partially screened by the proximity of the metal electrode.
For ferromagnetic tunneling contacts (i.e., tunnel barrier without any pinholes), since the voltage drop occurs across the tunnel barrier, one can study the bias induced polarization of the contacts\cite{19_Kamalakar2016_SciRep_Inversion_CVDhBNgr,249_gurram2017_NComms_biasInducedP}.

In order to bias the injector contact, we sweep DC current bias ($I_{\text{in}}$) along with a fixed amplitude of AC current $i$ = 1 $\mu$A. We use the standard lock-in detection technique to measure the voltage ($v$) across the non-local detector contacts, and obtain the non-local differential spin resistance $R_{\text{nl}}(I_{\text{in}}) = \frac{v}{i}$ at each value of the applied injection current bias $I_{\text{in}}$.
Figure~\ref{fig:Figure_5} shows the non-local differential spin signals $\Delta R_{\text{nl}} = (R_{\text{nl}}^{\text{P}} -R_{\text{nl}}^{\text{AP}})/2 $, measured at zero magnetic field, as a function of the bias applied across the injector contacts in Dev2 and Dev3.

For Dev2 with LR contacts, application of the current bias up to $\pm$50 $\mu$A (equivalent voltage bias, $V_{\text{in}}$ $\approx$ $\pm$0.07 V) across the injector resulted in a small change in $\Delta R_{\text{nl}}$ of around 0.06 $\Omega$[Fig.~\ref{fig:Figure_5}(a)].
The signal $\Delta R_{\text{nl}}$ is measured when the entire graphene channel is p-type at the carrier density n = -5$\times$10$^{12}$ cm$^{-2}$. 
Within the bias range of $\pm$0.07 V, the magnitude of $\Delta R_{\text{nl}}$ increases(decreases) with the positive(negative) bias.
Therefore, this behaviour could be due to transparent regions of the LR injector resulting in a finite voltage drop in the graphene leading to a strong local carrier drift underneath the metallic Co electrode\cite{31_Jozsa2009_PRB_ControlP_drift}.

For Dev3, Fig.~\ref{fig:Figure_5}(a) shows two sets (labelled, set1 and set2) of data for two different injector-detector contacts pairs. 
Each set consists of a HR injector and a LR detector. 
Under zero bias condition i.e., $V_{\text{in}}$ = 0, both sets show an inverted spin valve and Hanle signals $R_{\text{nl}}$ [shown in Figs.~\ref{fig:Figure_2}(d) and \ref{fig:Figure_2}(g) for set1]. 
When a DC current bias is applied across the injector contact up to $\pm$40 $\mu$A (equivalent voltage bias $V_{\text{in}}$ $\approx$ $\pm$0.3 V for set1, and $\approx$ $\pm$0.2 V for set2), the value of $\Delta R_{\text{nl}}$ changes in a peculiar way, which is independent of the gate voltage[Figs.~\ref{fig:Figure_5}(b) and \ref{fig:Figure_5}(c), for set1 and set2, respectively]. 
The sign of $\Delta R_{\text{nl}}$ remains same within the applied gate and bias range.
Interestingly, the magnitude of $\Delta R_{\text{nl}}$ increases at large negative bias.


The bias dependent spin signals $\Delta R_{\text{nl}}$ for Dev3 in Fig.~\ref{fig:Figure_5}(a)
are measured when the entire graphene channel is n-type. 
Within the bias range; $\pm$0.3 V for set1 and $\pm$0.2 V for set2, the magnitude of $\Delta R_{\text{nl}}$ increased(decreased) for the higher negative(positive) bias.
Moreover, we also measured the same behaviour when the carrier density of the graphene between the electrodes was changed to the vicinity of charge neutrality point and to the p-type, using the back gate voltage[Figs.~\ref{fig:Figure_5}(b)-\ref{fig:Figure_5}(c)]. 
These observations imply that the carrier density in graphene underneath the contact is screened by the metallic Co electrode due to possible transparent regions in the HR injectors of Dev3.
Due to HR nature of the injectors in Dev3, the voltage drop is mostly across the two-layer-CVD-hBN tunnel barrier. At a small bias range close to zero, we observe a peculiar behaviour of $\Delta R_{\text{nl}}$ which does not comply with the contact induced local carrier drift\cite{31_Jozsa2009_PRB_ControlP_drift}. 
We attribute this behaviour to bias induced spin polarization of the two-layer-CVD-hBN tunnel barrier.

We also measured Hanle spin signals $\Delta R_{\text{nl}}(B_{\text{z}})$ at different injection current biases for set1 and set2 contacts of Dev3.
Using the values of $\lambda_{\text{s}}$ obtained from the fitting of $\Delta R_{\text{nl}}(B_{\text{z}})$ data measured at different injection bias, we calculate $p_{\text{in}}$ of the (inner)injector contact using the following equation\cite{23_Tombros2007_Nat}:
\begin{equation}	\label{Eq_dRnl}
\Delta R_{\text{nl}} = p_{\text{in}} p_{\text{d}} \left( \frac{R_{\text{sq}} \lambda_{\text{s}} e^{-\frac{L}{\lambda_{\text{s}}}}}{2W} \right),
\end{equation}
where, $p_{\text{d}}$ is the differential spin detection polarization of the (inner)detector.
We assume that $p_{\text{d}}$ is constant, as the bias is applied only across the injector contact, and is equal to the unbiased $p_{\text{in}}$ of the injector, i.e., $p_{\text{d}}$ = $p_{\text{in}}(V_{\text{in}}=0)$.
The resulting $p_{\text{in}}$ at different injection bias voltages for the injectors in set1 and set2 are shown on the right y-axes of Figs.~\ref{fig:Figure_5}(b) and \ref{fig:Figure_5}(c), respectively. 
The change in $p_{\text{in}}$ as a function of bias nearly follows the change in $\Delta R_{\text{nl}}$, and the sign of $p_{\text{in}}$ remains negative. Moreover, the magnitude of both $\Delta R_{\text{nl}}$ and $p_{\text{in}}$ increases at higher negative bias, and the value of $p_{\text{in}}$ reaches up to -15\% at -0.3 V for set1, and at -0.2 V for set2 contacts of Dev3.

Kamalakar \textit{et al}.\cite{19_Kamalakar2016_SciRep_Inversion_CVDhBNgr} showed a similar inversion behaviour of spin signals for thicker (2-3 layers) CVD-hBN barriers over a large range of bias, $\pm$2 V, where the magnitude of the spin signal decreases at large injection bias voltages $|V_{\text{in}}| >$ 0.5 V. However, the authors\cite{19_Kamalakar2016_SciRep_Inversion_CVDhBNgr} do not report the data for smaller bias voltages $|V_{\text{in}}| <$ 0.5 V, the range within which we measure the differential spin signal $\Delta R_{\text{nl}}$ and differential spin polarization $p_{\text{in}}$ ($|V_{\text{in}}| < \pm$ 0.3 V).
Note that we used the low-frequency lock-in detection technique which helps to measure the spin signals even at a very small DC bias\cite{249_gurram2017_NComms_biasInducedP} which is difficult with the pure DC measurements\cite{19_Kamalakar2016_SciRep_Inversion_CVDhBNgr}.
On the other hand, recent report by Gurram \textit{et al}.\cite{249_gurram2017_NComms_biasInducedP} with bilayer-exfoliated-hBN tunnel barrier showed a dramatic change in $\Delta R_{\text{nl}}$ and $p_{\text{in}}$ with the applied bias, and their sign inversion near zero bias. We do not observe such inversion in sign of spin signals with bias for the two-layer-CVD-hBN barriers. This marks the different nature of the bilayer-exfoliated-hBN and two-layer-CVD-hBN tunnel barriers with respect to spin injection.

\section{Conclusions}
In conclusion, we have investigated room-temperature spin transport in graphene, encapsulated by a layer-by-layer-stacked two-layer-CVD-hBN tunnel barrier and a few-layer-thick exfoliated-hBN substrate.
Even though the graphene is supported by the bottom-hBN substrate, its mobility is quite low and thus resulted in small diffusion constants.
The lower values of mobilities and spin relaxation times compared to the already reported graphene on hBN devices are attributed to the conventional wet transfer technique used for transferring CVD-hBN tunnel barrier, and possible copper residues trapped in between the two CVD-hBN monolayers and at the interface with graphene.
We analyse the spin transport data by considering Elliott-Yafet and D'Yakonov-Perel' spin relaxation mechanisms and find no clear dominance of either of the mechanisms.

For the cobalt/two-layer-CVD-hBN/graphene/hBN contacts, we find no correlation between the $R_{\text{c}} A$ values of high-resistive contacts and the sign of the spin polarization.
Furthermore, spin polarization of the high-resistance contacts remains reversed with respect to the low-resistance contacts, within $\pm$0.3 V bias, and its magnitude increases at large negative bias. This behaviour is different from what has been reported for the contacts with high-resistive thick-CVD-hBN barriers, bilayer-exfoliated-hBN barriers, and oxide barriers.

We emphasize that the two-layer is different from the bilayer where the former is just an assembly of two individual monolayers and the latter is as-grown.
Despite having equivalent thicknesses, the two-layer-CVD-hBN barrier shows completely different bias dependence to that of the bilayer-exfoliated-hBN barrier\cite{249_gurram2017_NComms_biasInducedP}.
This implies that the quality and the relative alignment of two monolayers of hBN might play significant role in determining the tunneling characteristics.

We observe large magnitude of spin polarization up to 15\% at -0.2 V bias and it could be enhanced further with application of higher bias for high-resistance contacts with two-layer-CVD-hBN barriers which is promising for spintronics applications. 
However, in order to establish the role of CVD based hBN in graphene spintronics, it is important to prepare a clean device without hampering the quality of graphene for long distance spin transport. 
For this, recently proposed dry transfer technique for CVD grown materials\cite{580_drogeler2017_APL_dryTransfer_SV} could be adopted to greatly improve the quality of graphene spin valve devices.
Futhermore, we expect that a controlled growth of bilayer-CVD-hBN\cite{579_gao2013_ACSNano_Growth_1L2L_CVDhBN} tunnel barriers followed by dry transfer on top of a recently obtained high-quality CVD-graphene\cite{590_banszerus2015_SciAdv_CVDGr} could help to progress the role of CVD grown materials for spintronics in van der Waals heterostructures.


\bigbreak
We kindly acknowledge J.G. Holstein, H.M. de Roosz, H. Adema and T.J. Schouten for technical assistance. The research leading to these results has received funding from the European Union Seventh Framework Programme under Grant Agreement No. 696656 Graphene Flagship, Swiss Nanoscience Institute, Swiss National Science Foundation, iSpinText FlagERA project, National Natural Science Foundation of China (No. 51290272), and supported by the Zernike Institute for Advanced Materials and the Nederlandse Organisatie voor Wetenschappelijk (NWO).
  



%

\end{document}